# On Classification of MIMO Equalizers

Wing Chau Ng and Chuandong Li

Huawei Technologies, Ottawa, Canada, wing.chau.ng@huawei.com

**Abstract** *Four classes of MIMO equalizers are presented based on a DSP-perceived model, with the corresponding partial or full channel inverses. The complete channel inverse shows a dependency on the complex conjugates of equalizer inputs, coinciding with the widely linear equalization theory. A complete version was submitted to ECOC 2021.*

**Introduction**

A decade ago, digital coherent optical receivers were primarily designed to undo the optical channel impairments, such as dispersion compensation (CD), polarization rotation, polarization mode dispersion (PMD), laser frequency offset (FO) and phase noise (PN) etc.[1],[2]. These impairments take place on the optical fields, and therefore the digital signal processing (DSP) design was based on complex-valued inputs and complex-valued outputs, via adaptive equalizers using the ideal transmitted fields as references. As electrical impairments (such as s21 and phase mismatches, crosstalk) limit the next-generation coherent transceivers, it is indispensable to re-consider the overall channel starting from a four-dimensional (4D) real time-domain (TD) transmitted vector, $\vec{\iota}_t = (i_{X,t}, q_{X,t}, i_{Y,t}, q_{Y,t})^T \in \mathbb{R}_t^{4\times 1}$ to a 4D received vector, $\vec{r}_t = (r_{XI,t}, r_{XQ,t}, r_{YI,t}, r_{YQ,t})^T \in \mathbb{R}_t^{4\times 1}$. Instead of a real physical model, one should consider the channel based on what the DSP can "perceive".

In this theoretical work, the DSP-perceived channel is first simplified, based on which we categorize linear MIMO equalizers into four classes according to their reference locations. The entire channel inverse can be represented by a complex conjugate-dependent system, coinciding with the widely linear equalization theory[3]. Suboptimally removing FO dynamics, relatively static channel inverses parameterized with common device and channel parameters are presented for monitoring or calibration purposes.

**Channel perceived by DSP**

Assuming negligible nonlinearity, the 4D real electrical waveforms interact with linear channel components represented by their TD 4×4 real transfer matrices consisting of real responses with limited bandwidths (both DSP and devices have limited bandwidth). Following the chronological order of impairments, the 4D received electrical waveforms are

$$\vec{r}_t = D_{Rx,t} * R_{4\times 4}^{OE} * R_{4\times 4}^{RPN} R_{4\times 4}^{FO}(h_{4\times 4}^{CD} * U_{4\times 4}) * R_{4\times 4}^{TPN} T_{4\times 4}^{EO} * D_{Tx,t} * \vec{\iota}_t,$$

where $*$ denotes convolution. For simplicity, first, assume zero electrical crosstalks, $p=\{Tx,Rx\}$, $D_{p,t} = diag(D_{1,t}^p, D_{2,t}^p, D_{3,t}^p, D_{4,t}^p) \in \mathbb{R}_t^{4\times 4}$, where the frequency response of each electrical component, $D_{i,t}^p$, is $D_i^p(\omega) = |D_i^p(\omega)|e^{-j\theta_i^p(\omega)}$. Second, $T_{4\times 4}^{EO}, R_{4\times 4}^{OE} \in \mathbb{R}_t^{4\times 4}$ are the responses of IQ modulators and integrated coherent receiver (ICR), respectively. Third, $R_{4\times 4}^{pPN}, R_{4\times 4}^{FO} \in \mathbb{R}_t^{4\times 4}$ are PN at transmitter (Tx) and receiver (Rx) and time-varying (generally) FO, $\omega_{LO}(t)$, respectively, whose entries are real sinusoids in time:

$$R_{4\times 4}^{pPN} = diag(R_{2\times 2}^{pPN}, R_{2\times 2}^{pPN}) \in \mathbb{R}_t^{4\times 4}, R_{2\times 2}^{pPN} = \begin{bmatrix} \cos\theta_{pPN}(t) & -\sin\theta_{pPN}(t) \\ \sin\theta_{pPN}(t) & \cos\theta_{pPN}(t) \end{bmatrix} \in \mathbb{R}_t^{2\times 2};$$

$$R_{4\times4}^{FO} = diag(R_{2\times2}^{FO}, R_{2\times2}^{FO}) \in \mathbb{R}_t^{4\times4}, R_{2\times2}^{pPN} = \begin{bmatrix} \cos\omega_{LO}(t) & -\sin\omega_{LO}(t) \\ \sin\omega_{LO}(t) & \cos\omega_{LO}(t) \end{bmatrix} \in \mathbb{R}_t^{2\times2}.$$

Fourth, the bracket indicates that the optical signals experience distributive CD, $h_{4\times4}^{CD}$, and polarization effect, $U_{4\times4}$, where $U_{4\times4}$ is a unitary matrix,

$$U_{4\times4} = \begin{bmatrix} a & b & -c & d \\ -b & a & -d & -c \\ c & d & a & -b \\ -d & c & b & a \end{bmatrix},$$

which is parameterized by "nearly-static" $a, b, c, d \in \mathbb{R}$ where $a^2 + b^2 + c^2 + d^2 = 1$, i.e., $U_{4\times4} * \cong U_{4\times4}$.

Rx DSP algorithms are designed to equalize the channel in a "quasi-reverse" order of the impairments: Rx response and quadrature error compensation come first, and optical impairment equalization follows. FO compensation (C) is performed before CDC because $R_{4\times4}^{FO}$ and $h_{4\times4}^{CD}$ are not commutative (convolution and multiplication do not commute). Signal clocks need to be recovered first via CD compensation before a 2×2 complex MIMO reverses the polarization effect. $h_{4\times4}^{CD}$ thus seemingly appears after $U_{4\times4}$, but, for small PMD, CD and polarization effects are commutative, i.e., $h_{4\times4}^{CD} * U_{4\times4} \cong U_{4\times4} h_{4\times4}^{CD} *$. Carrier phase recovery cannot distinguish Rx PN from Tx PN, resulting in mis-ordered DSP stages and thus equalization enhanced phase noise (EEPN)[4], $R_{4\times4}^{EEPN}$, and the DSP perceives the channel as if

$$\vec{r}_t \cong D_{Rx,t} * R_{4\times4}^{OE} * R_{4\times4}^{FO} h_{4\times4}^{CD} * U_{4\times4} R_{4\times4}^{EEPN} T_{4\times4}^{EO} D_{Tx,t} * \vec{\iota}_t.$$

Finally, $T_{4\times4}^{EO} * D_{Tx,t} *$ are compensated by two 2×2 real MIMO equalizers[2] or 4×4 for XY crosstalks. $T_{4\times4}^{EO}$, $R_{4\times4}^{OE}$, are ignored as they can be absorbed into $D_{p,t}$ as crosstalks. Combine $R_{4\times4}^{PN}$ into $R_{4\times4}^{FO}$ as they share the same matrix form, the new model becomes

$$\vec{r}_t = D_{Rx,t} * R_{4\times4}^{FO}(h_{4\times4}^{CD} * U_{4\times4}) D_{Tx,t} * \vec{\iota}_t$$

in Fig. 1b or

$$\vec{r}_t = D_{Rx,t} * R_{4\times4}^{FO}(U_{4\times4} h_{4\times4}^{CD}) * D_{Tx,t} * \vec{\iota}_t, \tag{1}$$

With the commutativity between $R_{4\times4}^{FO}$ and $U_{4\times4}$ (provable by expanding), a back-to-back model becomes, shown in Fig. 1c,

$$\vec{r}_t = D_{Rx,t} * U_{4\times4} R_{4\times4}^{FO} D_{Tx,t} * \vec{\iota}_t. \tag{2}$$

In this paper, four MIMO equalizer classes will be defined based on their reference locations in Fig. 1d. Their functionalities will be discussed to match some previous DSP designs[2],[5-7] in parameter estimation or calibration.

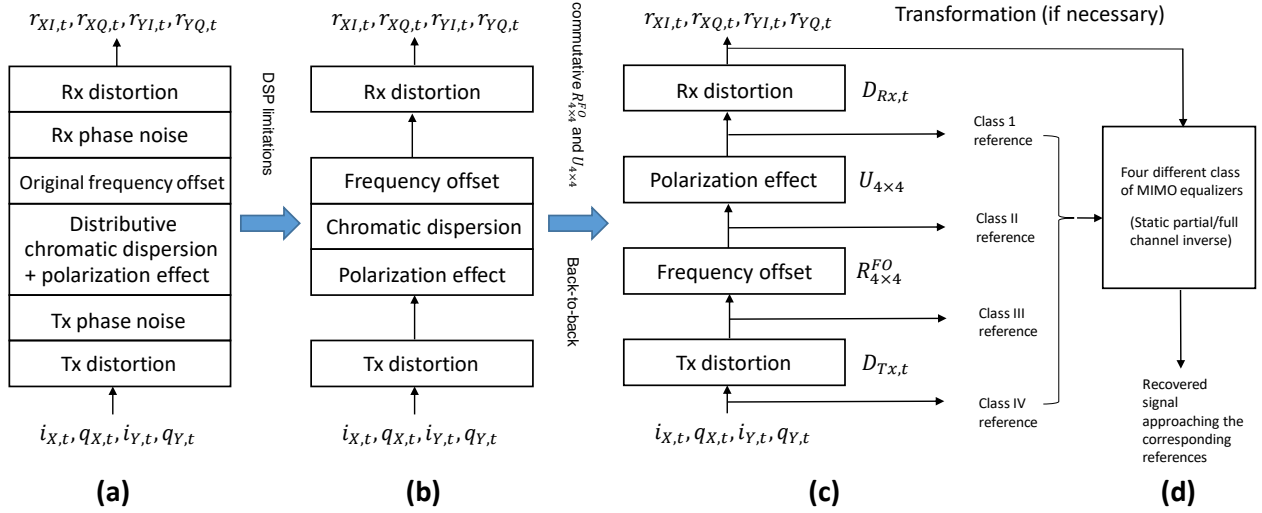

**Fig. 1:** (a) Channel model (b) Channel model perceived by Rx DSP (c) Simplified model with various reference locations for four equalizer classes (d) Overview of equalization.

The optimal Class II MIMO equalizer corresponds to the following channel inverse:

$$\mathbf{H}_{opt} \cong \begin{bmatrix} ae^{j\omega\tau_1^{Rx}} & -be^{j\omega\tau_2^{Rx}} & ce^{j\omega\tau_3^{Rx}} & -de^{j\omega\tau_4^{Rx}} \\ be^{j\omega\tau_1^{Rx}} & ae^{j\omega\tau_2^{Rx}} & de^{j\omega\tau_3^{Rx}} & ce^{j\omega\tau_4^{Rx}} \\ -ce^{j\omega\tau_1^{Rx}} & -de^{j\omega\tau_2^{Rx}} & ae^{j\omega\tau_3^{Rx}} & be^{j\omega\tau_4^{Rx}} \\ de^{j\omega\tau_1^{Rx}} & -ce^{j\omega\tau_2^{Rx}} & -be^{j\omega\tau_3^{Rx}} & ae^{j\omega\tau_4^{Rx}} \end{bmatrix}. \tag{3}$$

When the reference fields (X, Y) are used, the channel inverse can be expressed as, for some A and B,

$$\mathbf{H}_{2\times4,\text{opt}} \cong \begin{bmatrix} Ae^{j\omega\tau_1^{Rx}} & Ae^{j\omega\tau_2^{Rx}} & Be^{j\omega\tau_3^{Rx}} & Be^{j\omega\tau_4^{Rx}} \\ -B^*e^{j\omega\tau_1^{Rx}} & -B^*e^{j\omega\tau_2^{Rx}} & A^*e^{j\omega\tau_3^{Rx}} & A^*e^{j\omega\tau_4^{Rx}} \end{bmatrix}. \tag{5}$$

To move on, we employ "field matching" to FO-rotate the previous reference fields,

$$\begin{bmatrix} \tilde{s}_{X,o,t} \\ \tilde{s}_{Y,o,t} \end{bmatrix} = \mathbb{G}_{4R\to2C} \begin{bmatrix} s_{XI,t}\,e^{-j\omega_{LO}t} \\ s_{XQ,t}\,e^{-j\omega_{LO}t} \\ s_{YI,t}\,e^{-j\omega_{LO}t} \\ s_{YQ,t}e^{-j\omega_{LO}t} \end{bmatrix} = e^{-j\omega_{LO}t}\mathbb{G}_{4R\to2C}U_{4\times4}^{-1}D_{Rx,t}^{-1} * \vec{r}_t, \tag{6}$$

where the real-to-complex transformation matrix is

$$\mathbb{G}_{4R\to2C} = \begin{bmatrix} 1 & j & 0 & 0 \\ 0 & 0 & 1 & j \end{bmatrix} \in \mathbb{C}^{2\times4}.$$

Suboptimally, to avoid IQ mixing, one can first transform the received signal as follows:

$$\begin{bmatrix} \tilde{s}_{X,o,t} \\ \tilde{s}_{Y,o,t} \end{bmatrix} \cong \mathbb{G}_{4R\to 2C} U_{4\times 4}^{-1} D_{Rx,t}^{-1} * \begin{bmatrix} r_{XI}e^{-j\omega_{FO}t} \\ r_{XQ}e^{-j\omega_{FO}t} \\ r_{YI}e^{-j\omega_{FO}t}t \\ r_{YQ}e^{e^{-j\omega_{FO}t}} \end{bmatrix}. \tag{7}$$

Thus, the above allows us to reach class III MIMO equalizer, suboptimally having the same form of channel inverse as that of class II MIMO equalizer.

## Conclusions

Based on a "DSP-perceived" channel model, four classes of MIMO equalizers are presented according to their reference locations. Class I characterizes Rx response. Class II compensates Rx and polarization effects. Class III is class II free of FO dynamics, and it is extended to suboptimal class IV, enabling a complete, static channel inverse from Rx to Tx. Finally, a general channel inverse appears as a time-varying complex-conjugate dependent system which aligns with the widely linear equalization theory.

## Acknowledgement

The authors would express their gratitude to Huawei Technologies, Dongguan, Guangdong province, China for their massive experimental verification of each MIMO from March 2020 to August 2020. The authors would acknowledge the works of Rios-Müller et al.[3] ,and T. Kobayashi et al.[5]that brought insight to this theoretical work.